\documentclass[twocolumn,showpacs]{revtex4}

\usepackage{amsmath}

\usepackage{graphicx}

\begin{document}
\draft
\title{The low-frequency response
       in the surface superconducting state of ZrB$_{12}$ single crystal}

\author{Grigory I. Leviev, Valery M. Genkin, Menachem I. Tsindlekht, and Israel Felner}
\affiliation{The Racah Institute of Physics, The Hebrew University
of Jerusalem, 91904 Jerusalem, Israel}
\author{Yurii B. Paderno and  Vladimir B. Filippov}
\affiliation{Institute for Problems of Materials Science, National
Academy of Sciences of Ukraine, 03680 Kiev, Ukraine}

\begin{abstract}

The large nonlinear response of a single crystal ZrB$_{12}$ to an
ac field (frequency 40 - 2500 Hz) for $H_0>H_{c2}$ has been
observed. Direct measurements of the ac wave form and the exact
numerical solution of the Ginzburg-Landau equations, as well as
phenomenological relaxation equation, permit the study of the
surface superconducting states dynamics. It is shown, that the low
frequency response is defined by transitions between the
metastable superconducting states under the action of an ac field.
The relaxation rate which determines such transitions dynamics, is
found.

\end{abstract}

\date{\today}
\maketitle

\section{Introduction}

Recently high-quality superconducting ZrB$_{12}$ single crystals
with transition temperatures $T_c = 6.06$ K have been grown. The
investigation of their physical properties, including electron
transport, tunnel characteristics, and critical fields, have shown
that the Ginzburg-Landau parameter $\kappa$ is only slightly
larger than the boundary between the type-I - type-II
superconductor value~\cite{DAG,TS1}. In this paper we concern the
low-frequency response of a ZrB$_{12}$ crystal when the dc
external magnetic field $H_0>H_{c2}$ is parallel to the sample
surface. In spite of the fact that the sample is in the surface
superconducting state (SSS)~\cite{PG}, no static magnetic moment
is observed, while the ac response in this regime is large and
nonlinear even for an ac amplitude $h_0<<H_0$. Indeed, at
equilibrium the total surface current equals zero in SSS, the
internal dc magnetic field in the bulk equals $H_0$, and the
magnetic moment of the specimen is small. On the other hand, the
ac magnetic field drives the sample into a metastable SSS where
the total surface current reaches a finite value. The internal
magnetic field deviates from the external one, and as a results
the ac response becomes large. The low-frequency response of
superconductors in SSS was the focus of intensive experimental
investigations~\cite{ROLL,KAR,HOP} since the first prediction of
the existence of SSS in~\cite{PG}. The observed~\cite{ROLL} wave
form of the ac response, corresponding to the flux passing through
the specimen, explicitly invoked a model similar to the Bean model
~\cite{BEAN}. Recently SSS attracted renewed interest from various
directions as described in Ref.~\cite{TS2,TH,SARA,JUR,RYDH,FINK1}.
Paramagnetic effect in a superconducting disk~\cite{TH},
stochastic resonance~\cite{TS2}, , the percolation transition in
the field $H_0=0.81H_{c3}$~\cite{JUR} have been observed. It was
proposed to use low-frequency response for testing the quality of
superconducting resonators in accelerators~\cite{SARA}. Surface
states were observed also in single crystals of MgB2~\cite{RYDH}.
Our experimental results presented here show that in ZrB$_{12}$
single crystal, the Bean critical model of surface sheath does not
give an adequate description of the observed wave form which
corresponds to the flux passing through the sample. In the
framework of Ginzburg-Landau, theory we calculated the surface
current in metastable SSS's which exists under an ac magnetic
field. Observing the wave forms, we studied the metastable SSS
dynamics and determined the relaxation rate under an ac field. We
found that the relaxation time for transition to the equilibrium
state is not constant and depends on the surplus of the free
energy. This relaxation time is decreased with the dc magnetic
field, and depends on the driving field frequency.

\section{Experimental}

\begin{figure}
     \begin{center}
    \leavevmode
       \includegraphics[width=0.9\linewidth]{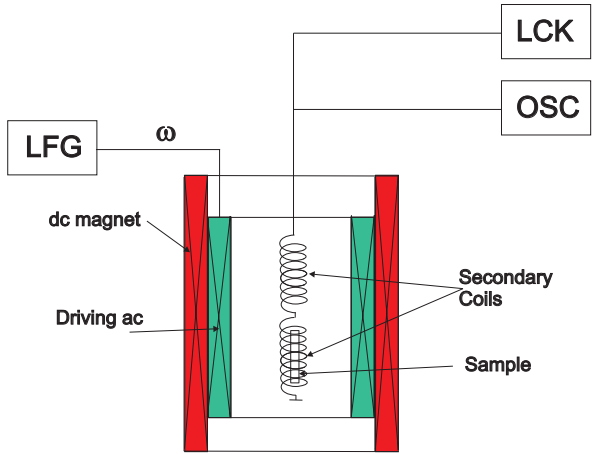}
       \bigskip
       \caption{(Color online) Block diagram of the experimental setup.
       LFG - low frequency generator,
        LCK - lock-in amplifier and OSC - oscillograph.}
     \label{f-1}
     \end{center}
     \end{figure}

The measurements were carried out at $T=5$ K° on ZrB$_{12}$ single
crystal. The sample was grown in the Institute for Problems of
Materials Science, Ukraine. Its dimensions are $10.3\times
3.2\times 1.2$ mm$^3$ and it was cut by an electric spark from a
large crystal of 6 mm diameter and 40 mm length. The surface of
the sample was polished mechanically and then chemical etched in
boiling HNO$_3$/H$_2$O (1:1) for 10 minutes was used. X-ray
pictures showed that a sample was single-phase material with the
UB$_{12}$ structure (space group Fm3m, $a=7.407$ \AA~\cite{KEN}.
The tunnel characteristics of this sample were described
earlier~\cite{TS1}. The dc-magnetic moment was measured using a
SQUID magnetometer. A block diagram of the ac linear and nonlinear
setup is shown in Fig.~\ref{f-1}.

The ac magnetic field $h(t)=h_0\sin(\omega t)$ was supplied by the
magnetometer copper solenoid. The ac response was measured by an
inductive pick-up coil method~\cite{SH}. The sample was put into
one coil of a balanced pair of pick-up coils and the induced
voltage $V(t)\propto dM(t)/dt$ was measured with an oscilloscope.
Here $M$ is the magnetic moment of the sample. The lock-in
amplifier was used in order to measure simultaneously in-phase and
out-of-phase signals of the first and third harmonics of the
driving frequency. An oscilloscope measured the wave form of the
signal in one channel. The second channel of the oscilloscope
measured the time derivative of the excitation field.

\section{Experimental results}

\begin{figure}
     \begin{center}
    \leavevmode
       \includegraphics[width=0.9\linewidth]{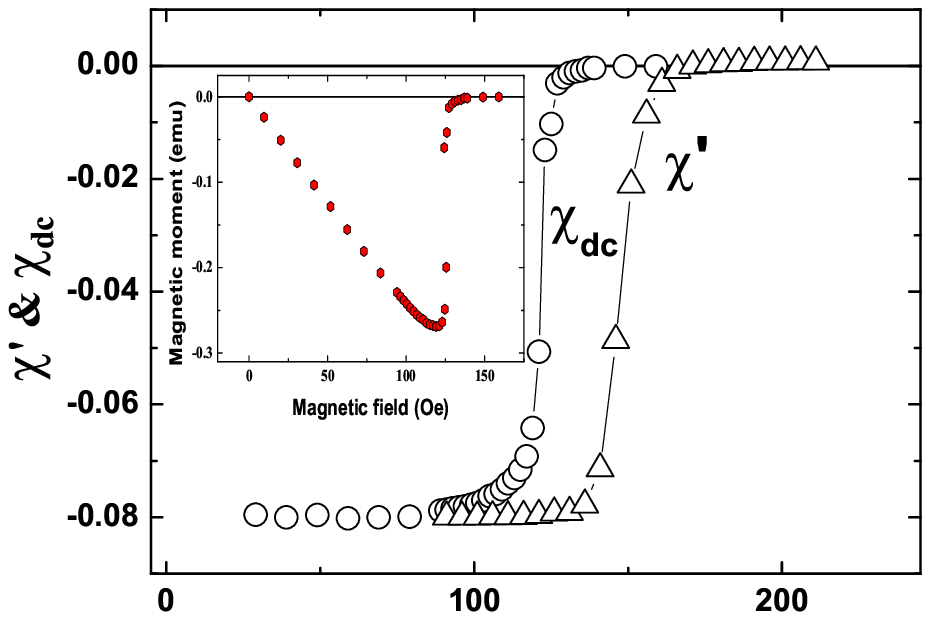}
       \includegraphics[width=0.9\linewidth]{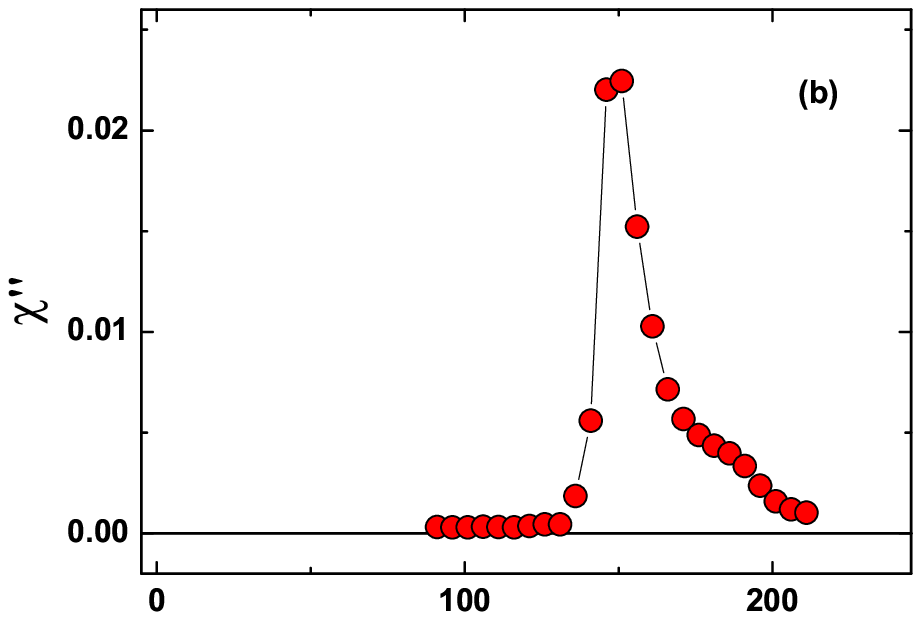}
       \includegraphics[width=0.9\linewidth]{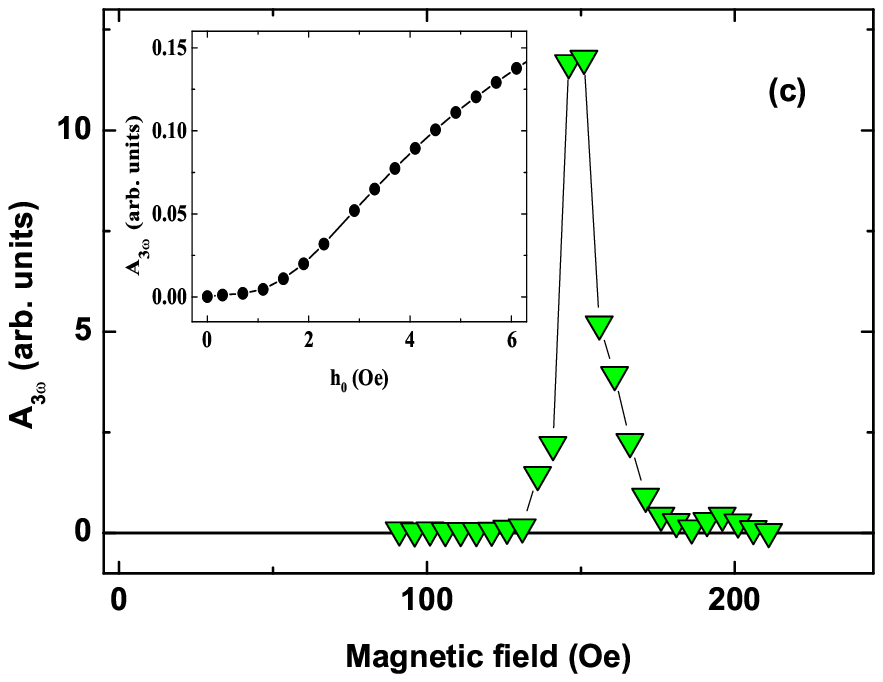}
       \bigskip
       \caption{(Color online) (a) $\chi{'}$ and $\chi_{dc}=M/H_0$
       magnetic field dependencies at $T=5$~K.
       Inset: magnetization curve after ZFC.\\
       (b) Magnetic field dependence of $\chi{''}$.\\
       (c) Field dependence of the amplitude of the third harmonic, $A_{3\omega}$.
        Inset: amplitude dependence of $A_{3\omega}(h_0)$ at $H_0=180$~Oe.\\
        ac measurements were carried out at frequency $\omega/2\pi=170$~Hz and $h_0=0.4$~Oe.}
     \label{f-2}
     \end{center}
     \end{figure}

Fig.~\ref{f-2}a shows the real part of the ac susceptibility at
the fundamental frequency, $\chi^{'}$, and the zero-field cooled
(ZFC) dc susceptibility, $\chi_{dc}=M/H_0$ as a function of dc
field $H_0$. The inset to Fig.~\ref{f-2}a presents the ZFC
magnetization curve at 5 K. Field dependencies of the ac
susceptibility imaginary part at fundamental frequency,
$\chi^{''}$, and amplitude of the third harmonic, $A_{3\omega}$
are shown on Fig.~\ref{f-2}b and Fig.~\ref{f-2}c respectively.
Amplitude dependence of the third harmonic, $A_{3\omega}(h_0)$ for
$H_0=180$~Oe is presented on inset to Fig.~\ref{f-2}c. It is clear
that this dependence is far to be cubic as the perturbation theory
predicts. Experiment shows that amplitude dependence of
$A_{3\omega}(h_0)$ is not cubic at any dc field $H_0$.

It is clear that the observed large signal of the $A_{3\omega}$
and maximum of the $\chi^{''}$ located in a magnetic field
$H_{c2}<H_0<H_{c3}$, e.g. in a surface superconducting state,
although the zero dc signal indicates that bulk of the sample is
in the normal state. The absorption in the SSS exceeds losses in
the mixed and normal states.

\begin{figure}
     \begin{center}
    \leavevmode
       \includegraphics[width=0.9\linewidth]{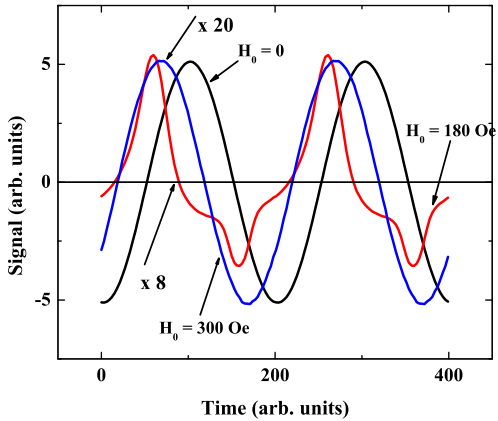}

       \bigskip
       \caption{(Color online) Oscillogram of $dM/dt$ for three different magnetic
       fields in the Meissner state ($H_0=0$), in surface superconducting state,
($H_0=180$~Oe~$<H_{c3}$),
    and for a normal state ($H_0=300$~Oe~$>H_{c3}$) at $h_0=4.75$~Oe.}
     \label{f-3}
     \end{center}
     \end{figure}
Fig.~\ref{f-3} shows the time derivative of the magnetic moment of
a sample at different applied magnetic fields at T = 5 K. Note (i)
that only in the SSS, the signal does not have the sine-form. (ii)
The amplitude dependence of the third harmonic,
$A_{3\omega}(h_0)$, presented in the inset to Fig.~\ref{f-2}c does
not exhibit any cubic dependence.

The experimental data presented in Figs.~\ref{f-2} and~\ref{f-3}
are complex and the theoretical model which explains these
observations is given in the next section.

\section{Theoretical model}

Our theoretical approach is based on the numerical solution of the
two Ginzburg-Landau equations~\cite{PDG} for the order parameter
and vector potential, which in the normalized form are as follows:

 \begin{equation}\label{Eq1}
  \begin{array}{c}
    -(i{\nabla}/\kappa+\overrightarrow{A})^2\Psi^2+\Psi-\mid\Psi\mid^2\Psi=0\\
   -\text{curl}\text{curl}\overrightarrow{A}=\overrightarrow{A}\mid\Psi\mid^2+ i/2\kappa(\Psi^*\nabla\Psi-\Psi\nabla\Psi^*)\\
  \end{array}
\end{equation}

The order parameter, $\Psi$, is normalized with respect to its
value in zero magnetic field, the distances with respect to the
London penetration length $\lambda$, and the vector potential
$\overrightarrow{A}$ with respect to $\sqrt 2 H_c\lambda$, where
$H_c$ is the thermodynamic critical field and $\kappa=\lambda/\xi$
is the Ginzburg-Landau parameter, and $\xi$ is the correlation
length. In the second equation for the vector potential we
neglected the normal current, assuming that the skin depth exceeds
both the London penetration length and sample thickness. It is
assumed that the sample form is a slab with 2d thickness and that
the external magnetic field is parallel to its surface. The chosen
coordinates are: the $x$-axis normal to the slab, (thus the
symmetry plane is $x = 0$), and the $z$-axis is directed along
magnetic field. It is assumed that external magnetic field
$H>H_{c2}=\kappa$.

Assuming the surface solutions have the form of

\begin{equation}\label{Eq2}
   \Psi(x,y) = f(x)exp(i\kappa ky)
\end{equation}
therefore, Eqs.~\ref{Eq1} reduce to
\begin{equation}\label{Eq3}
\begin{array}{c}
  -\frac{1}{\kappa}\frac{\partial^2 f}{\partial x^2 }+(A-k)^2f-f+f^3=0 \\
 \frac{\partial^2 A}{\partial x^2}=f^2(A-k),\\
\end{array}
\end{equation}
where $k$ is constant. The boundary conditions at $x=\pm d$ are:
$\partial f(\pm d)/\partial x =0$, $\partial A(\pm d)/\partial x =
H$ ; at $x=0$ $f(0)=0$ and A(0)=0; $H=H_0+h_0\text{sin}(\omega
t)$. An additional condition for the surface states $\partial
f(0)/\partial x = 0$ is satisfied only asymptotically for
$d\longrightarrow \infty$. For the equilibrium state the value of
$k=k_{eq}$ can be obtained by minimizing the Gibbs free energy
defined as:
 \begin{equation}\label{Eq4}
\begin{array}{c}
 \tilde{F}=\int dV\{\frac{1}{2}|\Psi|^4-|\Psi|^2+|i\nabla\Psi/\kappa +A\Psi|^2+ \\
+B^2-2BH\},
\end{array}
\end{equation}
where B is the magnetic induction. Using Eq.~\ref{Eq3} one then
obtains
\begin{equation}\label{Eq5}
\begin{array}{c}
  \tilde{F}=-HA(d)-\int^{d}_0
dx\{\frac{1}{2}f^4(x)+A(x)[A(x)-\\
-k]f^2(x)\}\\
\end{array}
\end{equation}

 The two coupled Eqs.~\ref{Eq3} could be solved by numerical methods.
The order parameter for surface solutions deviates from zero only
near the sample boundary, and we could consider comparatively
small $d/\lambda \leq 10$. The actual sample thickness exceeds
$\lambda$ by 4 or 5 orders of magnitude. The solutions for large
$d$ could be found from the ones for small $d$ by transformation
$k=k_s+H_i(d-d_s)$, where $H_i=H_s(0,H,K)$ is the magnetic field
at x = 0 in the problem for $d=d_s\cong 10\lambda$. The index $s$
corresponds to the solution for this small $d$. This choice of
$d_s$ is sufficient for numerical calculations and provides the
solutions with $f_s(0)=0$, $\partial f_s(0)/\partial x =0$. The
free energy transformed as

 \begin{equation}\label{Eq5a}
    \tilde{F}=\tilde{F_s}-H_i(d-d_s)(H-\int[A_s(x)-k_s]f^2_s(x)dx)
\end{equation}

\begin{figure}
     \begin{center}
    \leavevmode
       \includegraphics[width=0.9\linewidth]{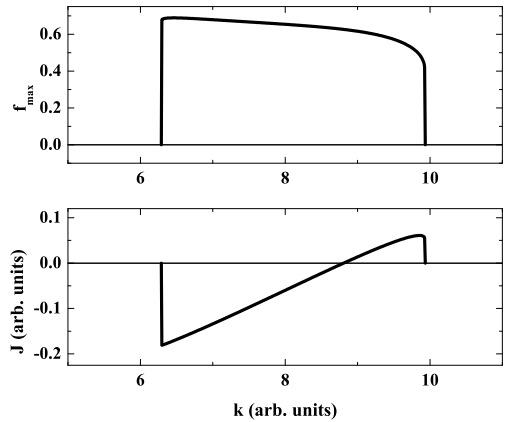}
       \bigskip
       \caption{Upper panel. The maximal value of the order parameter,
       $f_{max}$, as a function of $k$.
       Lower panel. Total surface current, $J$, as a function of the $k$ parameter.
        The equilibrium value of $k=8.8$ corresponds to zero surface current.}
     \label{f-4}
     \end{center}
     \end{figure}

\begin{figure}
     \begin{center}
    \leavevmode
       \includegraphics[width=0.9\linewidth]{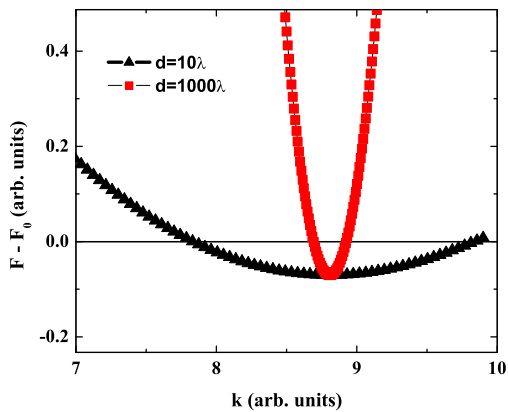}

       \bigskip
       \caption{(Color online) Free energy of the surface superconducting
       states relative to the free energy of normal state ($F_0$)
       in magnetic field $H_0=1$ and $k=0.75$ for two values of the sample thickness.}
     \label{f-5}
     \end{center}
     \end{figure}

To simplify the calculations we use below variable $k_s$ and omit
index $s$. The properties of the equilibrium solutions for a
semi-infinite half-space have been discussed in~\cite{FINK}.  The
order parameter, the supercurrent, and the internal magnetic field
were calculated. In these states the total surface current equal
zero and free energy reaches a minimum value. The ac magnetic
field drives the superconductor into a metastable state. These
states correspond to the solutions of Eqs.~\ref{Eq3} for $k\neq
k_{eq}$. The solution of Eqs.~\ref{Eq3} shows that surface states
exist in a wide range of $k$ near $k_{eq}$ as shown in the upper
panel of Fig.~\ref{f-4} , but only for $|k-k_{eq}|<<k_{eq}$ free
energy of these states is lower than the energy of the normal
state. Moreover, this range shrinks with increasing sample
thickness (Fig.~\ref{f-5}).

 This is due to the increase of the
contribution of the first term in Eq.~\ref{Eq5} which is the order
of the Gibbs energy of the normal state $\tilde{F_0}=-H^2d$. The
total surface current equals zero for equilibrium k and increases
with increasing $|k-k_{eq}|$  as is shown in the lower panel of
Fig.~\ref{f-4}. For a given $k$, the free energy versus magnetic
field does not exhibit any minimum in the equilibrium. Only the
difference between the Gibbs energy of the superconducting and
normal states exhibits minimum near equilibrium field as was
discussed in~\cite{ROLL}, but for a such representation the
reference point moved with changing field.

The magnetic moment of the sample actually depends on the total
surface current $J$, because the current is localized in a thin
surface layer. This current is a function of the external magnetic
field $H$ and $k$, $J=J(H,k)$. The response of the sample to the
ac magnetic field depends on the dynamics of $k$. A priori, one
can assume that the equation that governs the dynamics is

 \begin{equation}\label{Eq7}
    \frac{dk}{dt}=-\nu[k-k_{eq}(H)],
\end{equation}

where $k_{eq}(H)$ is the equilibrium $k$ in instantaneous value of
external magnetic field and $\nu = \nu(k-k_{eq})$ is the
relaxation rate. Function $k_{eq}(H)$ has to be found from
Eqs.~\ref{Eq3} and for $|H-H_0|<<H_0$ is well approximated by a
polynomial of the third order of $h=H-H_0$. Using the function
$J(h,k)$ calculated from Eqs.~\ref{Eq3} and Eq.~\ref{Eq7} one can
obtain the time evolution of the surface current in an ac field
and compare with observed wave forms. The time derivative of the
surface current is proportional to the observed signal $V$. The
coefficient $\alpha =\frac{1}{V}\frac{dJ}{dt}$ depends on the
experimental apparatus parameters. Actually we could obtain $k(t)$
directly from experimental data and test the correctness of
Eq.~\ref{Eq7}. We may write

\begin{equation}\label{Eq8}
 \frac{\partial J(h,k)}{\partial h}\frac{dh}{dt}+\frac{\partial J(h,k)}{\partial k}%
 \frac{dk}{dt}=\alpha V(t)
\end{equation}

This expression permits us to obtain $k(t)$ from the observed wave
form. It is a first order differential equation for $k(t)$. To
evaluate $k(t)$ we have to know $k$ at $t=0$ and $\alpha$, since
the derivatives $\partial J/\partial h$ and $\partial J/\partial
k$ can be calculated from Eqs.~\ref{Eq3}. The $k(0)$ value can be
found from the condition when the maximal current value during the
period is minimal. In order to find $\alpha$ we calculated $J(t)$
assuming that $\nu$ in Eq.~\ref{Eq7} is constant. Then, we choose
the $\nu$ value in order to minimize difference between the
calculated and experimental data. This procedure gives both $\nu$
and $\alpha$. To be sure that $\nu$ is actually constant, one has
to collect the weak ac field data. The observed signal during one
period of the ac field and the result of a simulation with
Eqs.~\ref{Eq3} and~\ref{Eq7} are shown in Fig.~\ref{f-4}. The data
in this figure were collected in a dc field of 130 Oe, and an ac
field with amplitude 1.78 Oe and frequency $\omega/2\pi= 733$ Hz.
In our calculations we took $\nu/\omega  = 0.05$ and the
Ginzburg-Landau parameter $\kappa = 0.75$~\cite{TS1}. The good
correlation between the calculated and experimental data permits
one to find the scale coefficient $\alpha$ which is used below.

\section{discussion}

\begin{figure}
     \begin{center}
    \leavevmode
       \includegraphics[width=0.9\linewidth]{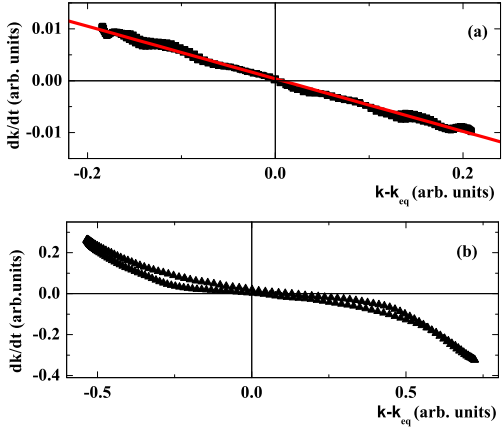}

       \bigskip

\caption{(Color online) The time derivative $dk/dt$ plotted as a
function of $k-k_{eq}$ for at $H_0=130$~Oe and $\omega/2\pi
=733$~Hz (a): $h_0=1.78$~Oe and (b): $h_0=5.9$~Oe. The linear fit
of the experimental curve (a) gives $\nu/\omega=0.051$. The
hysteresis at large amplitudes (b) shows that, generally speaking,
$dk/dt$ depends on $k$ and the instantaneous magnetic field not
only through $k=k_{eq}$.}
     \label{f-7}
     \end{center}
     \end{figure}

As was shown above the losses are small both in the mixed and
normal states and have a maximum at $H_0 > H_{c2}$ (see
Fig.~\ref{f-2}a).  The $H_{c2}=126$~Oe is determined from the dc
magnetization curve (inset to Fig.~\ref{f-2}a). The oscillogram,
Fig.~\ref{f-3}, in both the Meissner and normal states ($H_0 = 0$
and 300 Oe) has a sine shape, and for $H_{c2} < H_0 < H_{c3}$ the
wave form deviates from a sine shape. We do not observe any clear
plateau for $dM/dt$ in the ac period. Such a plateau is a
peculiarity of the Bean model when it is applied to surface
currents~\cite{ROLL,FINK2}. Using the experimental data and the
model developed in previous section one can calculate $dk/dt$ as a
function of $k-k_{eq}$. Fig.~\ref{f-7} shows $dk/dt$ plotted as a
function of $k-k_{eq}$ obtained from the wave forms that were
observed for $H_0 = 130$ Oe and $\omega/2\pi = 733$ Hz. The linear
fit of $dk/dt$ at $h_0=1.78$ Oe yields $\nu(0)/\omega =0.051$
which agrees well with the $\nu/\omega =0.05$ used in
Eq.~\ref{Eq7} when the scale coefficient has been found. Visible
hysteresis in Fig.~\ref{f-7}b indicates that at a high amplitude
of excitation the relaxation rate in Eq.~\ref{Eq7} $\nu$ depends
on $k$ and on the instantaneous value of
 $h(t)$ not only through $k-k_{eq}(h)$.

\begin{figure}
     \begin{center}
    \leavevmode
       \includegraphics[width=0.9\linewidth]{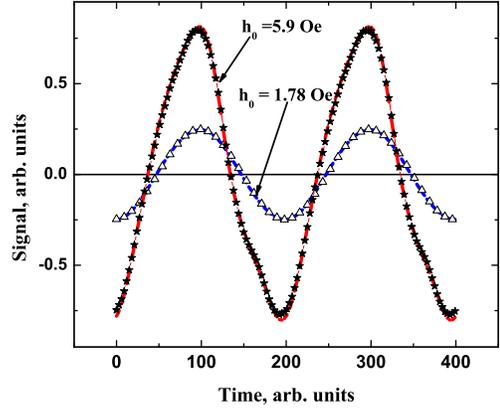}

       \bigskip
\caption{(Color online) The observed and calculated (solid lines)
oscillogram for $H_0=130$~Oe, $\omega/2\pi=733$~Hz at
$h_0=1.78$~Oe and $h_0=5.9$~Oe.}

 \label{f-8}
 \end{center}
 \end{figure}

The expression for $\nu(k-k_{eq})$ could be found from fitting of
$dk/dt$ by the polynomial of $k-k_{eq}$. The approximation
expression which has a form

\begin{equation}\label{Eq9}
\begin{array}{c}
  \nu(x)/\omega =0.051-0.117x+\\
+1.323x^2+0.184x^3-0.747x^4\\
\end{array}
\end{equation}

provides the calculated data, which with an accuracy of better
than 10\%, reproduces the experimental data for a dc field of
130~Oe and a frequency of 733 Hz as is shown in Fig.~\ref{f-8}.

\begin{figure}
     \begin{center}
    \leavevmode
       \includegraphics[width=0.9\linewidth]{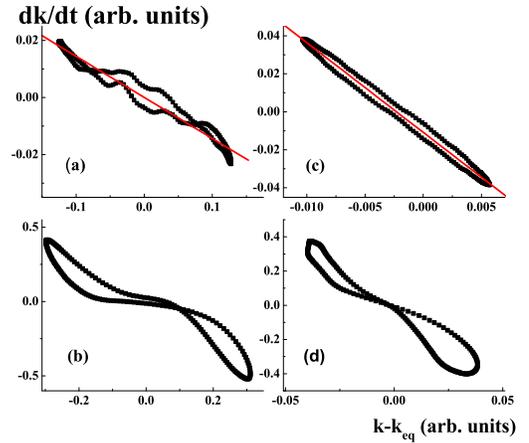}

       \bigskip
\caption{(Color online)$dk/dt$ as a function of $k-k_{eq}$ at
$\omega/2\pi=733$~Hz. (a) - $H_0=138$~Oe, $h_0=0.59$~Oe; (b) -
$H_0=138$~Oe, $h_0=5.9$~Oe; (c) - $H_0=180$~Oe, $h_0=0.59$~Oe; (d)
- $H_0=180$~Oe, $h_0=5.9$~Oe. Linear fit at low amplitude of
excitation ( (a) and (c) panels) gives $\nu/\omega=0.144$ for
$H_0=138$~Oe and $\nu/\omega=4.73$ for $H_0=180$~Oe.} \label{f-9}
     \end{center}
     \end{figure}

 Increasing the dc field
leads to the increasing of the relaxation rate $\nu$. We found
that for $H_0= 138$ and 180 Oe, the relaxation parameter for weak
ac amplitudes are $\nu(0)/\omega =0.145$ and 4.725 respectively
(see Fig.~\ref{f-9}).

\begin{figure}
     \begin{center}
    \leavevmode
       \includegraphics[width=0.9\linewidth]{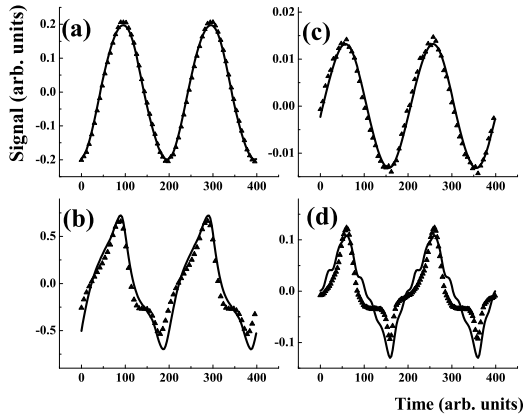}

       \bigskip
\caption{Calculated (solid lines) and experimental (triangle)
oscillograms at $\omega/2\pi=733$~Hz. (a) - $H_0=138$~Oe,
$h_0=0.59$~Oe; (b) - $H_0=138$~Oe, $h_0=5.9$~Oe; (c) -
$H_0=180$~Oe, $h_0=0.59$~Oe; (d) - $H_0=180$~Oe, $h_0=5.9$~Oe.}
     \label{f-10}
     \end{center}
     \end{figure}

The calculated wave forms with the help of the proposed model
reproduce experimental data only for a weak ac field as shown in
Fig.~\ref{f-10}. This is due to the increase in the difference
between the two values of $dk/dt$ for the same $k-k_{eq}$ at
larger ac amplitudes, ( see at Fig.~\ref{f-9}b, d).

\begin{figure}
     \begin{center}
    \leavevmode
       \includegraphics[width=0.9\linewidth]{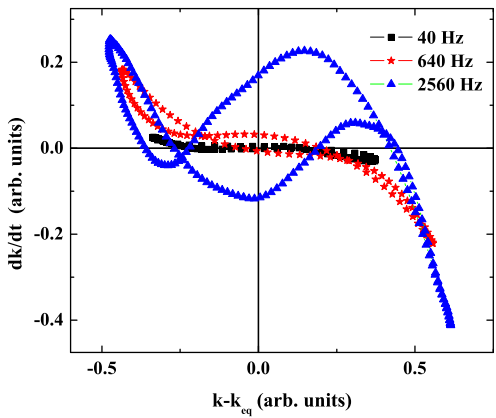}
       \bigskip
       \caption{(Color online) $dk/dt$ as a function of $k-k_{eq}$ at
       $H_0=130$~Oe and $h_0=4.75$~Oe for different frequencies.}
     \label{f-11}
     \end{center}
     \end{figure}

Fig.~\ref{f-11} shows the $dk/dt(k-k_{eq})$ dependence for
different frequencies at $H_0=130$~Oe and $h_0=4.75$~Oe. One may
conclude from this figure that the relaxation rate $\nu$ (if
Eq.~\ref{Eq7} could be applied) increases with excitation
frequency $\omega$.

\begin{figure}
     \begin{center}
    \leavevmode
       \includegraphics[width=0.9\linewidth]{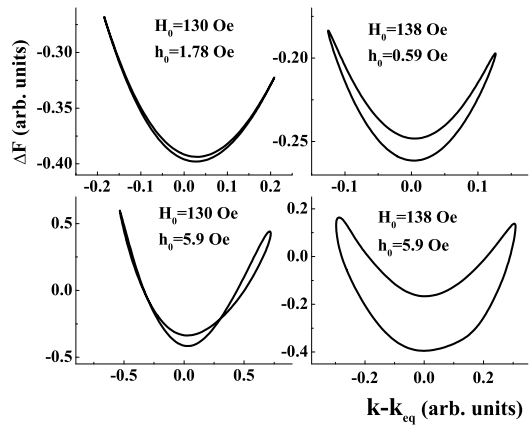}

       \bigskip
\caption{Calculated dependence of the Gibbs energy as a function
of the $k-k_{eq}$ parameter during the ac cycle for different
values of the dc field $H_0$ and amplitude of excitation, $h_0$ at
frequency $\omega/2\pi=733$~Hz.}
     \label{f-13}
     \end{center}
     \end{figure}

It is clear that the model equation~\ref{Eq7}, where the
relaxation rate $\nu$ depends on the one variable $k-k_{eq}$, is
valid only for small ac amplitudes. The transition from the
surface state with one $k$ to another $k$, requires changing the
order parameter in the whole sample. Possibly, it happens through
the nucleation  of a new phase. This process is governed by excess
energy. In general the relaxation constant in Eq.~\ref{Eq7} may
depend on the energy of the state. Approximately it could be taken
into account by the assumption that the relaxation constant
depends on $k-k_{eq}$. We see that for small ac fields and not far
away from $H_{c2}$, this assumption is correct. But increasing the
ac amplitude and/or the dc field results in the explicit
dependence on both: $k$ and the instantaneous value of the
magnetic field $H_0$. The straightforward calculations of the
Gibbs energy $F$, exhibited in Fig.~\ref{f-13}, shows that when
this energy is a single-valued function of $k-k_{eq}$, the
simulation with Eq.~\ref{Eq7} gives acceptable results.

When $F$ becomes a multivalued function, the calculated wave
forms differ from the experimental data. In order to obtain a proper
theoretical description, one has to take into account that the
energy of SSS is not expressed only through $k-k_{eq}$.

\section{Conclusion}

We experimentally investigated the dynamics of the surface
metastable superconducting states of ZrB$_{12}$ in ac fields at
low frequencies (40 - 2500 Hz). It was shown that for low ac
amplitudes of excitation these dynamics are governed by a simple
relaxation equation. The relaxation rate depends on the deviation
from the equilibrium state. Decreasing of the frequency of the
applied ac field results in increasing the relaxation time.

\section{Acknowledgments}

This work was supported by the INTAS program under the project No.
2001-0617.


\bigskip

\begin{references}

\bibitem{DAG} D. Daghero, R.S. Gonnelli, G.A. Ummarino, A. Calzolari,
Valeria Dellarocca, V.A. Stepanov, V.B. Filippov and Y.B. Paderno,
Supercod. Sci. Technol., {\bf 17}, S250 (2004).

\bibitem{TS1}  M.I. Tsindlekht, G.I. Leviev, I. Asulin, A. Sharoni, O. Millo,
 I. Felner, Yu.B. Paderno, V.B. Filippov, and M.A.
 Belogolovskii, Phys. Rev. B {\bf 69}, 212508 (2004)

\bibitem{PG} D. Saint-James and  P.G. Gennes, Phys. Lett. {\bf 7}, 306 (1963).

\bibitem{ROLL} R.W. Rollins and J. Silcox, Phys. Rev. 155, 404 (1967);
R.W. Rollins, R.L. Cappelletti, and J.H. Fearday, Phys. Rev. B
{\bf 2}, 105 (1970).

\bibitem{KAR} V.R. Karasik, N.G. Vasil'ev, and V.S. Vysotskii,
Sov. Phys. JETP {\bf 35}, 945 (1972), [Zh. Eksp. Teor. Fiz. {\bf
62}, 1818 (1972)].

\bibitem{HOP} J.R. Hopkins, D.K. Finnemore, Phys. Rev. B {\bf 9}, 108  (1974).

\bibitem{BEAN} C. Bean, Rev. Mod. Phys. {\bf 36}, 41 (1964).

\bibitem{TH} D.J. Thompson, M.S.M. Minhaj, L.E. Wenger, and J.T. Chen,
 Phys. Rev. Lett. {\bf 75}, 529 (1995); P. Kostic, B. Veal,
 A. P. Paulikas, U. Welp, V. R. Todt,
 C.Gu, U. Geiser, J.M. Williams, K.D. Carlson, and R.A. Klemm,
 Phys. Rev. B {\bf 53}, 791 (1996).

\bibitem{TS2} M.I. Tsindlekht, I. Felner, M. Gitterman, B.Ya. Shapiro,
Phys. Rev. B, {\bf 62}, 4073 (2000).

\bibitem{JUR} J. K\"{o}tzler, L. von Sawilski, and S. Casalbuoni, Phys. Rev. Lett.
 {\bf 92}, 067005-1 (2004).

\bibitem{SARA}  S. Casalbuoni, L. von Sawilski, and J. K\"{o}tzler,  cond-mat/0310565

\bibitem{RYDH} A. Rydh, U. Welp, J.M. Hiller, A.E. Koshelev, W.K. Kwok, G.W. Crabtree,
K.H. P. Kim, K.H. Kim, C.U. Jung, H.-S. Lee, B. Kang, and S.-I.
Lee, Phys. Rev. B {\bf 68}, 172502 (2003).

\bibitem{FINK1} H.J. Fink and S.B. Haley, Int. J. Mod. Phys. B {\bf 17}, 2171 (2003),
cond-mat/0303121.

\bibitem{KEN} C.H.L. Kennard and L. Davis, J. Solid State Chem. {\bf 47}, 103
(1983).

\bibitem{SH} D. Shoenberg, {\it Magnetic oscillations in metals},
(Cambridge University Press, Cambridge, 1984).

\bibitem{FINK} H.J. Fink, R.D. Kessinger, Phys. Rev. {\bf 140}, A1937 (1965).


\bibitem{FINK2} H.J. Fink, L.J. Barnes, Phys. Rev. Lett. {\bf 15}, 792
(1965).

\bibitem{PDG}  P.G. de Gennes, {\it Superconductivity of Metals and Alloys }(%
W.A.Benjamin, New-York, 1966).


\end{references}
\end{document}